\documentclass[openacc]{rstransa}%%%%where rstrans is the template name

\usepackage{gensymb}
\usepackage[export]{adjustbox}
\usepackage{lineno}

\makeatletter
\def\makeLineNumberLeft{%
	\linenumberfont\llap{\hb@xt@\linenumberwidth{\LineNumber\hss}\hskip\linenumbersep}% left line number
	\hskip\columnwidth% skip over column of text
	\rlap{\hskip\linenumbersep\hb@xt@\linenumberwidth{\hss\LineNumber}}\hss}% right line number
\leftlinenumbers% Re-issue [left] option
\makeatother

\begin{document}

%%%% Article title to be placed here
\title{Modelling H$_{3}^{+}$ in planetary atmospheres: effects of vertical gradients on observed quantities}

\author{%%%% Author details
L. Moore$^{1}$, H. Melin$^{2}$, J. O'Donoghue$^{3}$, \\T. Stallard$^{2}$, J. Moses$^{4}$, M. Galand$^{5}$, \\S. Miller$^{6}$ and C. Schmidt$^{1}$}

%%%%%%%%% Insert author address here
\address{$^{1}$Boston University, USA.\\
$^{2}$University of Leicester, UK. \\
$^{3}$NASA Goddard Space Flight Center, USA.\\
$^{4}$Space Science Institute, USA. \\
$^{5}$Imperial College London, UK. \\
$^{6}$University College London, UK.}
%%\address{$^{1}$First author address\\
%%	$^{2}$Second author address\\
%%	$^{3}$Third author address}

%%%% Subject entries to be placed here %%%%
\subject{Aeronomy, Giant Planets, Giant Planets}

%%%% Keyword entries to be placed here %%%%
\keywords{Aeronomy, Ionosphere, H$_{3}^{+}$}

%%%% Insert corresponding author and its email address}
\corres{L. Moore\\
\email{moore@bu.edu}}

%%%% Abstract text to be placed here %%%%%%%%%%%%
\begin{abstract}
Since its discovery in the aurorae of Jupiter \textasciitilde30 years ago, the H$_{3}^{+}$ ion has served as an invaluable probe of giant planet upper atmospheres.  However, the vast majority of monitoring of planetary H$_{3}^{+}$ radiation has followed from observations that rely on deriving parameters from column-integrated paths through the emitting layer.  Here, we investigate the effects of density and temperature gradients along such paths on the measured H$_{3}^{+}$ spectrum and its resulting interpretation.  In a non-isothermal atmosphere, H$_{3}^{+}$ column densities retrieved from such observations are found to represent a lower limit, reduced by 20\% or more from the true atmospheric value.  Global simulations of Uranus' ionosphere reveal that measured H$_{3}^{+}$ temperature variations are often attributable to well-understood solar zenith angle effects rather than indications of real atmospheric variability.  Finally, based on these insights, a preliminary method of deriving vertical temperature structure is demonstrated at Jupiter using model reproductions of electron density and H$_{3}^{+}$ measurements.  The sheer diversity and uncertainty of conditions in planetary atmospheres prohibits this work from providing blanket quantitative correction factors; nonetheless, we illustrate a few simple ways in which the already formidable utility of H$_{3}^{+}$ observations in understanding planetary atmospheres can be enhanced.
\end{abstract}
%%%%%%%%%%%%%%%%%%%%%%%%%%%

%%%%%%%%%% Insert the texts which can accomdate on firstpage in the tag "fmtext" %%%%%

\begin{fmtext}
\end{fmtext}

%%%%%%%%%%%%%%% End of first page %%%%%%%%%%%%%%%%%%%%%

\maketitle

%%\linenumbers    %%%% comment out if not using linenumbers in the doc

\section{Introduction}
%%%% Insert A head here
Hydrogen, the most abundant cosmic element, also dominates the composition of giant planets.  Consequently, the most prominent ion species in giant planet atmospheres are the stable H$^{+}$ and H$_{3}^{+}$ ions \cite{Tao2011}.  The proton, H$^{+}$, is not spectroscopically observable, whereas the spectrum of H$_{3}^{+}$ is exceptionally rich, particularly the $\nu_{2}$ vibration rotation band in the near-infrared \cite{Oka2013}, a spectral region accessible from Earth's high-altitude observatories.  In fact, the first astronomical spectroscopic detection of H$_{3}^{+}$, enabled by a confluence of evolving theoretical, laboratory, and observational advances, was made in Jupiter's auroral region \cite{Drossart1989}.  Further detections at Saturn and Uranus, and continued ground-based monitoring of H$_{3}^{+}$ at Jupiter, Saturn and Uranus in the subsequent decades, have demonstrated its remarkable effectiveness as a probe of giant planet upper atmospheres (e.g., \cite{Uno2014,Stallard2018,Melin2013,Melin2018,ODonoghue2018,Johnson2018,Dinelli2017,Galand2011}; and references therein).

At present, the field of comparative aeronomy -- that is, the comparative study of planetary upper atmospheres -- relies on a sparse sampling of remote diagnostics, especially for the giant planets.  Vertical thermal structure, in particular, is difficult to determine remotely, and yet it plays a primary role in identifying the relevant physical processes at work in planetary upper atmospheres.  There are only a handful of temperature profiles obtained for Jupiter \cite{Seiff1998}, Uranus and Neptune, primarily from the Voyager spacecraft \cite{Festou1981,Herbert1987,Broadfoot1989}, whereas Saturn has now been relatively thoroughly sampled by Cassini \cite{Koskinen2015,Koskinen2018,Yelle2018}.  

The observed exospheric temperatures at all of the giant planets are hundreds of Kelvin hotter than predictions based on solar heating alone, emphasizing a rather fundamental lack of understanding in the energy balance in giant planet atmospheres, and highlighting the need for more thorough spatiotemporal thermospheric temperature constraints.  Modellers are actively seeking an explanation for this energy discrepancy, which may simply involve redistribution of auroral energy inputs \cite{Muller-Wodarg2018,Yates2018,Majeed2005}, or perhaps alternative energy sources, such as wave-driven heating from below \cite{ODonoghue2016}.  In the meantime, measurements of H$_{3}^{+}$ temperatures are vital for bridging this knowledge gap, as H$_{3}^{+}$ is thought to be in quasi-LTE with the surrounding neutral atmosphere \cite{Melin2005,Miller2010,Tao2011}, and valuable insights have already been provided by H$_{3}^{+}$ observations to-date.  However, derived H$_{3}^{+}$ temperatures also suffer from a key ambiguity: the vast majority of ground-based observations are column integrations through the entire ionosphere, from top-to-bottom, and therefore a convolution of the vertical structures in both H$_{3}^{+}$ density and temperature.  

Here, we investigate how giant planet atmospheric models can help supplement interpretation of H$_{3}^{+}$ spectroscopic observations.  These calculations are concentrated on unravelling the column-integrated density and temperature degeneracy behind the observed spectra, and on minimizing -- or at least understanding -- the effect of thermospheric gradients on analysis of H$_{3}^{+}$ datasets.  First, in section \ref{Methods}, we briefly describe the modelling approach as well as the observational constraints used.  Next, in section \ref{Results and discussion}, we consider a series of increasingly realistic ionospheric models and examine the complications that can arise in interpreting column-averaged observations.  Finally, we combine the insights from these sections in order to demonstrate a preliminary method of retrieving altitude profiles of H$_{3}^{+}$ temperature from nadir viewing geometry observations.

\section{Methods} \label{Methods}
\subsection{Modelling overview} \label{Modelling overview}
%%%% Insert B head here
The majority of H$_{3}^{+}$ ions in giant planet ionospheres are in photochemical equilibrium (PCE), as the H$_{3}^{+}$ chemical lifetime is much shorter than the transport timescale at most altitudes, and therefore the ion continuity equation simplifies to equating local production and loss (i.e., \textit{P$_{s}$ = L$_{s}$}) \cite{Moore2018a,Schunk2009}.  While transport processes are still relevant -- and highly so for H$^{+}$, especially at high altitude -- the dominance of chemical loss at low altitudes justifies the use of one-dimensional (1-D) simulations over those regions, which offer the advantages of simplicity and computational freedom over more dynamically comprehensive 3-D simulations.

Ionization fractions at the giant planets are roughly of order 10$^{-6}$ \cite{Majeed2004}, indicating that ion chemistry and dynamics are largely inconsequential for the underlying neutral atmosphere.  However, this generality is less true at auroral latitudes, where there appears to be a correspondence between auroral emission signatures with complex hydrocarbons and stratospheric hazes \cite{Guerlet2015,Sinclair2017}, and where H$_{3}^{+}$ can act as a thermostat, helping to maintain a cooler thermosphere \cite{Miller2013} and limit atmospheric escape.

Two models are adopted in the present work, which is focused on non-auroral latitudes at Jupiter and Uranus.  The simulations are conducted in 1-D, owing to the prevalence of PCE for H$_{3}^{+}$ distributions, and there are separate neutral and plasma modules in order to enable a more computationally-efficient exploration of ion chemistry.

The neutral module is described in detail by \textit{Moses and Poppe }\cite{Moses2017}, and is actually a combination of a meteoroid ablation code \cite{Moses1992,Moses1997} with the Caltech/JPL 1-D KINETICS photochemical model \cite{Allen1981,Yung1984}.  KINETICS solves the coupled mass-continuity equations as a function of pressure, and (for the neutral species) includes molecular and eddy diffusion transport terms.  It has proved to be effective and highly adaptable, having been applied to all of the giant planets, and currently treats 70 hydrocarbon and oxygen species that interact via \textasciitilde500 recently-updated chemical reactions \cite{Moses2017,Moses2015}.  Input for the meteoroid ablation code follows from revised constraints on interplanetary dust fluxes in the outer Solar System based on in situ spacecraft data \cite{Poppe2016}.  The resulting oxygenated and hydrocarbon mixing ratios are in agreement with a wide range of observational constraints \cite{Moses2017}.  Therefore, after adjusting the KINETICS simulations for the solar and geometric conditions explored here, the resulting neutral atmospheres serve as an excellent background for exploring realistic ion-neutral photochemistry at the giant planets.

Plasma densities and temperatures follow from another 1-D model called BU1DIM (the Boston University 1-D Ionosphere Model).  BU1DIM was originally developed for Saturn \cite{Moore2004,Moore2006a}, though has since been applied to Earth \cite{Moore2006b} and Mars \cite{Matta2014}.  Its most recent iteration has been generalized for application to any planetary atmosphere, and includes significantly expanded chemistry \cite{Moore2018a}.  BU1DIM describes the time- and altitude-dependent structure of an ionosphere by solving the coupled continuity, momentum and energy equations for all ion species of interest.  Jupiter's magnetic field is specified using results from the Juno spacecraft \cite{Connerney2018}.  At Uranus, however, magnetic field measurements are limited to a single flyby \cite{Ness1986}.  The primary effect of  magnetic fields on 1-D ionospheric calculations is to constrain the plasma motion (e.g., introducing a sin$^{2}$\textit{I} term into the expression for vertical ion drift velocity, where \textit{I} is the magnetic dip angle \cite{Moore2004,Rishbeth1969}).  Therefore -- partly due to incomplete knowledge of Uranus' magnetic field, and partly due to the predominance of PCE at H$_{3}^{+}$ altitudes -- magnetic field lines at Uranus are considered to be vertical here in order to focus investigations on the effect of vertical thermospheric gradients on derived H$_{3}^{+}$ parameters.  Modelled ion production rates follow from the attenuation of solar Extreme UltraViolet (EUV; 10-121 nm) and soft X-ray photons (combined, the XUV) \cite{Galand2009}, which are extrapolated to Jupiter and Uranus based on measurements from the Thermosphere Ionosphere Mesosphere Energetics and Dynamics Solar EUV Experiment (TIMED/SEE) \cite{Woods2005}.  In addition, secondary ionization and thermal electron heating rates are specified using parameterizations derived from coupled electron transport calculations at Saturn \cite{Moore2009}.  Aside from from solar XUV radiation, no other sources of energy input are considered here (e.g., energetic particle precipitation).

Early theoretical models of giant planet ionospheres predicted electron densities that were up to an order of magnitude too large based on later spacecraft measurements, with Saturn exhibiting the most extreme discrepancy \cite{Waite1987,Majeed2004}.  One commonly adopted mechanism for reducing modelled electron densities in order to better match observations was to convert H$^{+}$ into a molecular ion via the reaction
%%% Numbered equation
\begin{align}\label{1.1}
%	\begin{split}
		\textit{H$^{+}$ + H$_{2}$($\nu$ $\ge$ 4) $\rightarrow$ H$_{2}^{+}$ + H}
%	\end{split}
\end{align}
Without the introduction of some form of ion-neutral charge-exchange reaction, such as \eqref{1.1}, modelled  H$^{+}$ -- and hence electron density, \textit{n$_{e}$} -- is unrealistically large, as the radiative recombination rate coefficient for  H$^{+}$ is extremely slow (\textasciitilde10$^{-12}$ cm$^{3}$ s$^{-1}$ for typical giant planet thermospheric electron temperatures) \cite{Kim1994}.  The \eqref{1.1} reaction rate is thought to be near its maximum kinetic value \cite{Huestis2008b,Huestis2008a}, however the fraction of molecular hydrogen in the 4$^{th}$ or higher vibrational state is not constrained by observations at present.  For Jupiter, we adopt the vibrational density results from calculations by \textit{Majeed et al.} \cite{Majeed1991a}, which lead to an effective H$_{2}$ vibrational rate coefficient in combination with \cite{Huestis2008a}.  For Uranus, as two of the dominant sources of vibrationally excited H$_{2}$ have been shown to be photon-induced fluorescence and dissociative recombination of H$_{3}^{+}$ ions \cite{Majeed1991a} -- two solar-driven processes -- we scale the fractional H$_{2}$ vibrational populations for Jupiter by \( \frac{1}{r^{2}} \) to account for the diminution of solar photons with distance.  Thus adjusted, the \textit{Majeed et al. }results are then interpolated onto the appropriate Uranus pressure grid.  Further model inputs and specific settings are discussed in relation to their corresponding results in section \ref{Results and discussion}.

\subsection{Observations and data reduction} \label{Observations and data reduction}
%%\subsubsection{Observations and data reduction}
%%%% Insert C head here
While this is primarily a modelling study, there are two primary sources of data used to constrain the model results at Jupiter.  First, the Galileo G0N radio occultation, obtained on 8 December 1995, sampled Jupiter's dusk ionosphere near $24\degree$ S latitude and $292\degree$ E longitude \cite{Yelle2004,Hinson1997}.  This measurement provides a representative ionospheric electron density profile suitable for demonstrating the effect of vertical temperature gradients on retrieved H$_{3}^{+}$ parameters, at least when combined with model simulations that reproduce both the Galileo electron density profile and subsequent H$_{3}^{+}$ column density observations.  

Such H$_{3}^{+}$ column densities are the second source of data in this study: ground-based, high-resolution spectroscopic observations of Jupiter in the L telluric window.  Jupiter high-resolution (R \textasciitilde 25,000) spectroscopic data spanning 3.26-4 $\mu$m were obtained over four nights in April 2016 (14$^{th}$, 16$^{th}$, 20$^{th}$, and 23$^{rd}$) using the Near InfraRed Spectrograph (NIRSPEC \cite{Mclean1998}) on the 10 m Keck II telescope, as described in \textit{Moore et al. }\cite{Moore2017b}.  Combined, these observations yielded global coverage of Jupiter, with overlapping data from 2+ nights for more than half of the planet.  For the current work, a spectrum corresponding to the G0N latitude and local time was extracted and reduced as described in \textit{Moore et al. }\cite{Moore2017b}.  Briefly, using the line list of \textit{Neale et al.} \cite{Neale1996} and the partition function and total emission formulation of \textit{Miller et al.} \cite{Miller2010}, assuming conditions of q-LTE, Gaussian line-fitting techniques \cite{Stallard2002} are used to fit the modelled spectra to observed H$_{3}^{+}$ R- and Q-branch lines in order to retrieve column-averaged vibrational temperatures and column-integrated densities \cite{Melin2014}.

\section{Results and discussion} \label{Results and discussion}
The following subsections are dedicated to investigating the errors introduced in retrieved H$_{3}^{+}$ densities and temperatures due to realistic vertical atmospheric gradients.  These errors are expected because standard reduction of H$_{3}^{+}$ observations, which typically have nadir viewing geometries, starts by assuming a uniform layer of constant density and temperature, whereas we know that neither density nor temperature are constant within giant planet H$_{3}^{+}$ layers.  First, in section \ref{Results and discussion}\ref{Effect of vertical atmospheric gradients on H$_{3}^{+}$ observables: Density}, we examine a range of realistic temperature and density gradients in order to quantify the degree to which retrieved H$_{3}^{+}$ column densities are underestimated.  Next, in \ref{Results and discussion}\ref{Effect of vertical atmospheric gradients on H$_{3}^{+}$ observables: Temperature} we produce global simulations of Uranus' ionosphere to demonstrate the effect of gradients on retrieved H$_{3}^{+}$ temperatures.  Finally, in \ref{Results and discussion}\ref{Vertical structure}, we combine H$_{3}^{+}$ spectral observations with electron density constraints to produce a realistic H$_{3}^{+}$ density profile at Jupiter, and we utilize the insights from \ref{Results and discussion}\ref{Effect of vertical atmospheric gradients on H$_{3}^{+}$ observables: Density} and \ref{Results and discussion}\ref{Effect of vertical atmospheric gradients on H$_{3}^{+}$ observables: Temperature} in order to extract a preliminary H$_{3}^{+}$ temperature profile.

\subsection{Effect of vertical atmospheric gradients: H$_{3}^{+}$ density} \label{Effect of vertical atmospheric gradients on H$_{3}^{+}$ observables: Density}
We first examine the effect of vertical gradients using a series of simplified, synthetic slab atmospheres.  A column-integrated line-of-sight will transect the H$_{3}^{+}$ layer in a planetary ionosphere at some oblique angle -- or along the zenith for observations from directly overhead.  Giant planet H$_{3}^{+}$ emissions are optically thin \cite{Lam1997}, and lie in a spectral region with strong methane absorption (near 3.4 $\mu$m) \cite{Connerney2000}.  The intensity of the H$_{3}^{+}$ spectrum in each slab scales linearly with density and exponentially with temperature \cite{Miller2013}.  Furthermore, no self-absorption is considered between slabs.  Thus, the observed spectrum follows from the net sum of each individual "slab" emission element within the ionosphere, where the slab emission depends on the column density and temperature of a slab.  The retrieved H$_{3}^{+}$ "temperature" will only represent the "true" atmospheric temperature for an infinitely thin or isothermal H$_{3}^{+}$ layer.  In all other, more realistic cases, the derived temperature and density are thus necessarily a result of the convolution of the altitude structure in both density and temperature.  

\begin{figure}[hbt!]
	\centering
	\includegraphics[width=0.8\linewidth]{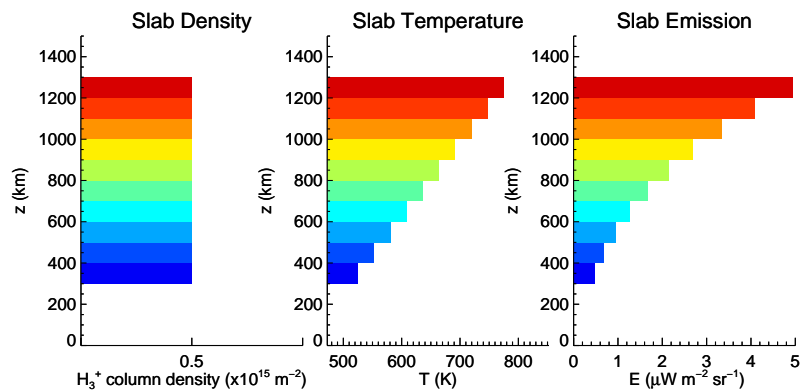}
	\caption{A simplified representation of Jupiter's H$_{3}^{+}$ layer based on \textit{Maurellis and Cravens }\cite{Maurellis2001}.  }
	\label{fig:figure-1}
\end{figure}

Based on previous modelling results \cite{Maurellis2001,Barrow2012}, we begin by approximating Jupiter's H$_{3}^{+}$ layer to have an equivalent slab width of \textasciitilde1000 km in altitude, as the calculated H$_{3}^{+}$ density is a nearly-constant \textasciitilde5000 cm$^{-3}$ over this altitude range \cite{Maurellis2001}.  This leads to an implied column density of \textasciitilde0.5x10$^{16}$ m$^{-2}$.  Over this same region, Jupiter's temperature increases from -- very approximately -- 525 K to 775 K \cite{Seiff1998}.  The temperature gradient is thus roughly 0.25 \( \frac{K}{km} \), and the mean temperature of the H$_{3}^{+}$ layer, \textit{T$_{mean}$}, is \textasciitilde650 K.  We then divide this synthetic H$_{3}^{+}$ layer into an arbitrary number \textit{n} slabs, each with column density \textit{N$_{slab}$} = 0.5x10$^{16}$ / \textit{n}, a temperature \textit{T$_{slab}$} based on the temperature gradient, and a corresponding slab emission.  We choose \textit{n} = 10 for this example, as fewer slabs are better-represented graphically, but discuss other variations below.  This general atmospheric structure is illustrated in Figure \ref{fig:figure-1}.
%\begin{figure}[hbt!]
%	\centering
%	\includegraphics[width=0.8\linewidth]{"../../../slab saves/Figure 2 - retrieved slab spectra"}
%	\caption{Synthetic slab spectra, color-coded according to the slab densities and temperatures in Figure \ref{fig:figure-1}.  The sum of the individual slab spectra is given by the black line while the dashed red line, which lies directly on top of the black one, is the retrieved H$_{3}^{+}$ model spectrum for N = 0.4x10$^{16}$ m$^{-2}$ and \textit{T} = 694.6 K.  Note that the spectral fit follows the standard assumption of a single layer at uniform temperature. }
%	\label{fig:figure-2}
%\end{figure}
\\ \indent The composite H$_{3}^{+}$ spectrum based on the structure from Figure \ref{fig:figure-1} will clearly dominated by higher altitude (redder, hotter) slab spectra.  By fitting this spectrum as described in \ref{Methods}\ref{Observations and data reduction}, we retrieve a column density and column-integrated temperature of 0.4x10$^{16}$ m$^{-2}$ and \linebreak695 K, respectively.  These values demonstrate an important complication with using observed H$_{3}^{+}$ spectra to derive realistic column densities, as the retrieved density from the simulated spectral fit is only 80\% of the true density.  This discrepancy is due to the exponential relation between H$_{3}^{+}$ temperature and emission, combined with the standard assumption that the "H$_{3}^{+}$ layer" is isothermal, and this weighting is also reflected in the retrieved temperature, 695 K, which is \textasciitilde7\% higher than the true mean temperature represented in \ref{fig:figure-1} (650 K).

A more realistic treatment of the simplified situation depicted above would complicate matters further due to the problems common to all astronomical observations, such as contaminations due to other emissions and absorptions, detector noise, and so forth.  Before stressing this initial \linebreak \( \frac{N_{fit}}{N_{true}} \) = 0.8 result too strongly, however, it is important to investigate the various sensitivities that lead to such a discrepancy in the column density.  For example, while the conditions in Figure \ref{fig:figure-1} have been chosen to roughly represent Jupiter's ionosphere, the mean temperature and the temperature gradient will be different at other locations at Jupiter and at other planets.  Similarly, we might question what effect a more realistic density profile would have.  The rest of this subsection is therefore devoted to outlining how different choices in representing the atmosphere and in generating the synthetic spectra affect the retrieved H$_{3}^{+}$ column density.

\begin{figure}[hbt!]
	\centering
	\begin{minipage}[t]{5cm}
		\centering
		\includegraphics[width=1.15\linewidth,center]{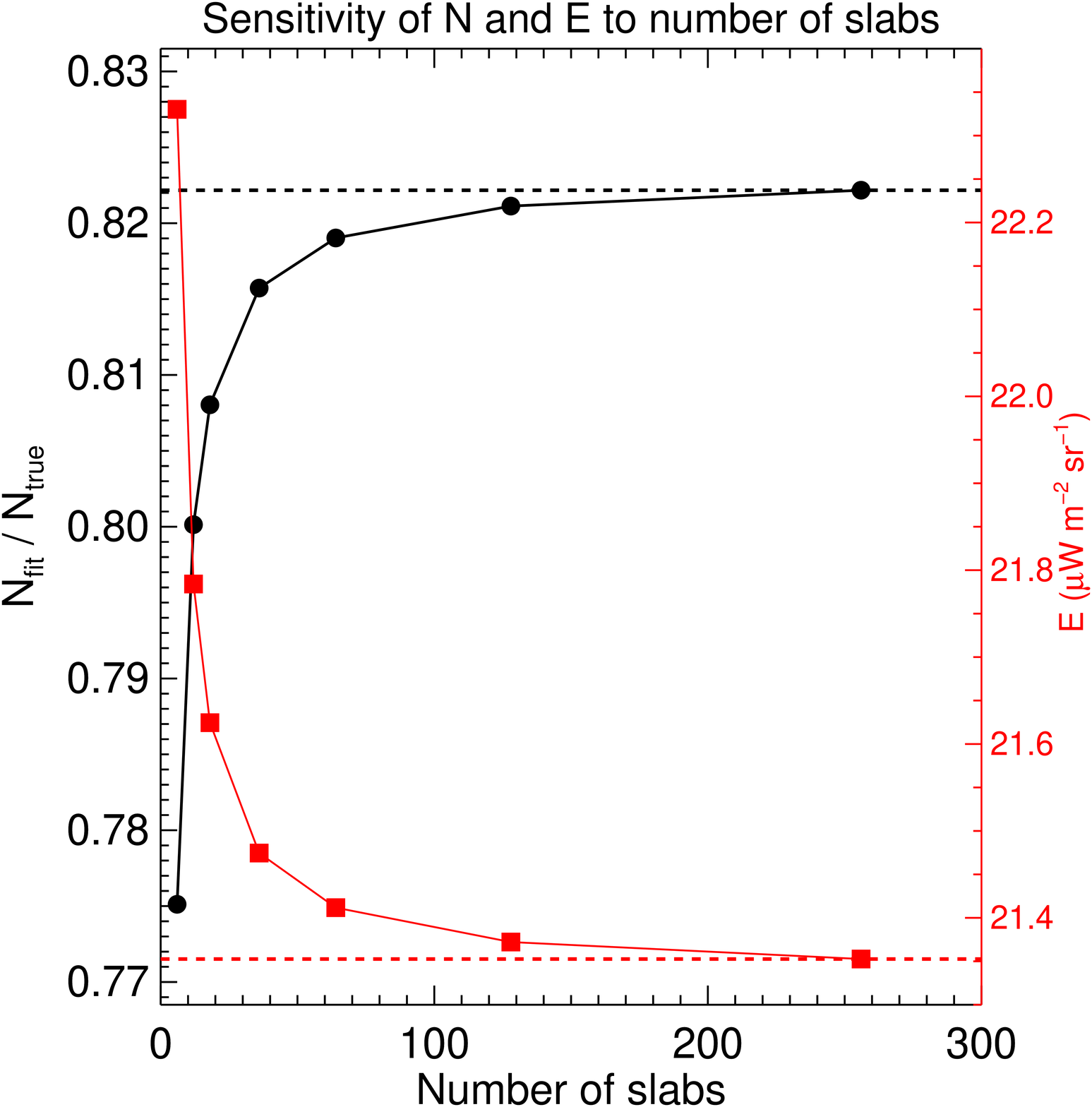}
		\caption{Total H$_{3}^{+}$ emission (red squares) and retrievals of H$_{3}^{+}$ column density \textit{N$_{fit}$}, relative to the true H$_{3}^{+}$ column density \textit{N$_{true}$} (black circles), based on slab spectra for the conditions introduced in Figure \ref{fig:figure-1}.  Dashed lines represents asymptotic values for large slab numbers.}
		\label{fig:figure-3}
	\end{minipage}
	\hspace{2cm}
	\begin{minipage}[t]{5cm}
		\centering
		\includegraphics[width=1.15\linewidth,right]{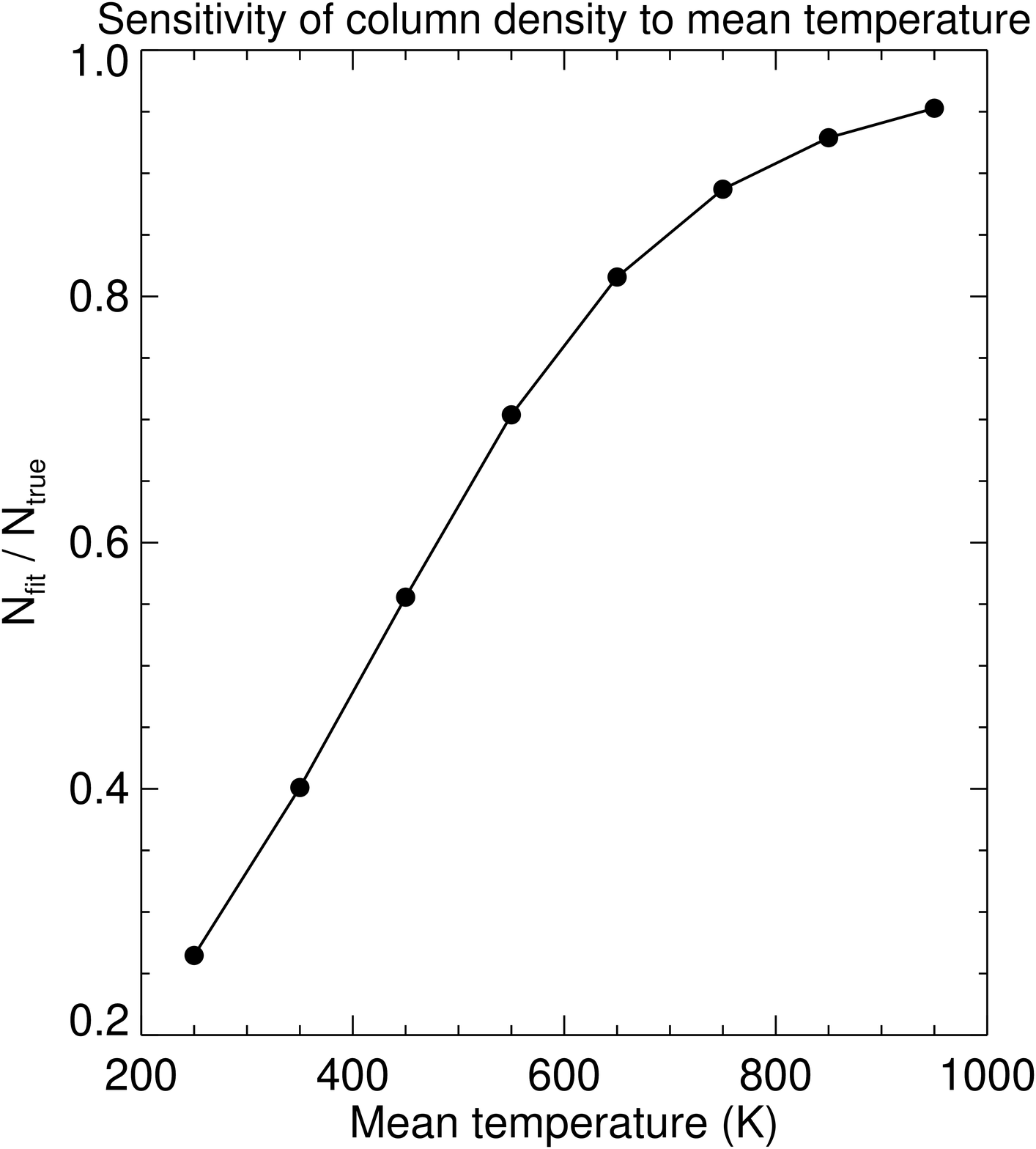}
		\caption{Retrievals of H$_{3}^{+}$ column density \textit{N$_{fit}$}, relative to the true H$_{3}^{+}$ column density \textit{N$_{true}$}.  The number of slabs \textit{n} is 36, and the mean temperature is varied while holding the vertical temperature gradient constant (i.e., as shown in Figure \ref{fig:figure-1}, 0.25 \( \frac{K}{km} \)).}
		\label{fig:figure-4}
	\end{minipage}
\end{figure}

First, we examine the choice of the number of slabs \textit{n} on the derived column density.  As demonstrated in Figure \ref{fig:figure-3}, increasing the number of slabs leads to an asymptotic approach towards \( \frac{N_{fit}}{N_{true}} \) = 0.822, which is a \textasciitilde3\% increase over the 10 slab representation imagined in Figure \ref{fig:figure-1}.  There is a corresponding inverted trend in the total H$_{3}^{+}$ emission with slab number, as representations with fewer slabs associate a wider range of the H$_{3}^{+}$ layer with higher temperatures.  In the example considered for Figure \ref{fig:figure-3}, the net emission decreases by \textasciitilde5\% towards an asymptotic value of \linebreak 21.35 $\mu$W m$^{2}$ sr$^{-1}$.  For comparison, typical giant planet models blanket the ionosphere with fewer than 100 grid points.  The derived column density will also depend on the mean temperature in the atmosphere, however.  Figure \ref{fig:figure-4} explores this effect while keeping the temperature gradient and the number of slabs fixed (i.e., the temperature gradient is as shown in Figure \ref{fig:figure-1}, 0.25 \( \frac{K}{km} \), and the mean temperature is varied).
\\ \indent Based on Figures \ref{fig:figure-3} and \ref{fig:figure-4}, derived H$_{3}^{+}$ column density is highly dependent upon the atmospheric temperature profile in the H$_{3}^{+}$ layer.  Before exploring a wider range of temperature gradients, such as might be more widely representative of H$_{3}^{+}$ in giant planet ionospheres, we investigate the additional impact of density gradients.  (It should be noted that the \( \frac{\textit{N$_{fit}$}}{\textit{N$_{true}$}} \) ratio is not sensitive to the true slab column density, which is perhaps not surprising given the linear relation between density and emission and the fractional error introduced from temperature gradients.)  Whereas the temperature profile is generally monotonically increasing in the lower thermosphere, there will be both positive and negative gradients in H$_{3}^{+}$ density.

\begin{figure}[hbt!]
	\centering
	\includegraphics[width=1.0\linewidth]{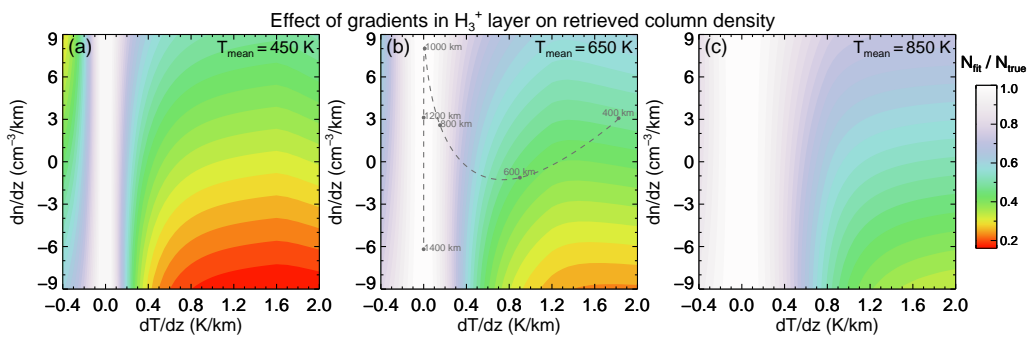}
	\caption{Contours of retrieved H$_{3}^{+}$ column density \textit{N$_{fit}$}, relative to the true H$_{3}^{+}$ column density \textit{N$_{true}$}, as a function of temperature and density gradients within the H$_{3}^{+}$ layer.  These results follow the approach outlined in Figure \ref{fig:figure-1}.  Specifically, an H$_{3}^{+}$ layer is divided into \textit{n} = 36 slabs, with a true slab column density of 0.5x10$^{16}$ m$^{-2}$ and with mean temperatures of \textbf{(a)} 450 K, \textbf{(b)} 650 K, and \textbf{(c)} 850 K.  Within those constraints, synthetic model spectra are then generated based on imposed density and temperature gradients, and \textit{N$_{fit}$} is derived from the composite synthetic spectrum.  The dashed line in panel (b) is described below in \ref{Results and discussion}\ref{Vertical structure}.}
	\label{fig:figure-5}
\end{figure}

Figure \ref{fig:figure-5} presents the combined effects of both density and temperature gradients in the H$_{3}^{+}$ layer on simulated retrieved column densities, \textit{N$_{fit}$}.  These results follow from an atmospheric layer divided into \textit{n} = 36 slabs, with a true slab column density \textit{N$_{true}$}=0.5x10$^{16}$ m$^{-2}$ and mean temperatures \textit{T$_{mean}$} of (a) 450 K, (b) 650 K, and (c) 850 K.  These calculations follow the approach outlined in Figure \ref{fig:figure-1}, except the absolute H$_{3}^{+}$ temperatures are allowed to range only between \linebreak 150-1250 K, temperatures appropriate for giant planet upper atmospheres.  This requirement means that, for the largest temperature gradients explored, there are regions of the H$_{3}^{+}$ layer that are isothermal (i.e., at either 150 K or 1250 K). As expected due to the linear relationship between H$_{3}^{+}$ emission and density, there is no discrepancy between modelled and "observed" column density when there is no temperature gradient.  The generally vertical contours in Figure \ref{fig:figure-5} indicate that \( \frac{\textit{N$_{fit}$}}{\textit{N$_{true}$}} \) is most sensitive to temperature gradients, though density gradients begin to play an important role when the temperature gradient is larger than \textasciitilde0.5 \( \frac{K}{km} \).  For realistic temperature gradients at the giant planets, \textasciitilde0.4-2 \( \frac{K}{km} \) \cite{Seiff1998,Yelle2018}, the retrieved H$_{3}^{+}$ column density ranges from \textasciitilde20-90\% of the true value, depending primarily on the mean temperature within the H$_{3}^{+}$ layer.  Observed uncertainties in \textit{N$_{fit}$} are often comparable, so the effect illustrated in Figure \ref{fig:figure-5} might be absorbed into experimental errors for low S/N data.  For instance, in a study of the anticorrelation between H$_{3}^{+}$ density and temperature, \textit{Melin et al.} \cite{Melin2014} find that a S/N of 12 is required to achieve a 10\% column density uncertainty for a column-averaged H$_{3}^{+}$ temperature of 600 K.
\\ \indent A more realistic H$_{3}^{+}$ layer would experience variations in both temperature and density gradients with altitude, and so could not be represented as simply as in Figure \ref{fig:figure-5}.  Careful examination of the evolution of dT/dz and dn/dz with altitude could give some idea of the net error induced in \textit{N$_{fit}$} based on those gradients, though such an analysis would also rely on prior knowledge of the atmosphere, and so significantly reduce the value of added observation.  Therefore, while the preceding figures establish that, unless the H$_{3}^{+}$ layer is in an isothermal atmosphere, the H$_{3}^{+}$ column density retrieved from observations will represent a lower limit, it is not practical at this stage to examine an all-inclusive range of possible atmospheric structures in order to assign definitive quantitative values to those lower limits.  Instead, these results serve as further motivation for investigating other similar complications of interpreting H$_{3}^{+}$ observations, and for outlining "toy" model parameters that would be relevant for development of a full H$_{3}^{+}$ retrieval model (e.g., \cite{Irwin2008}).

\subsection{Effect of vertical atmospheric gradients: H$_{3}^{+}$ temperature (Uranus)} 
\begin{figure}[hbt!]
	\centering
	\includegraphics[width=0.75\linewidth]{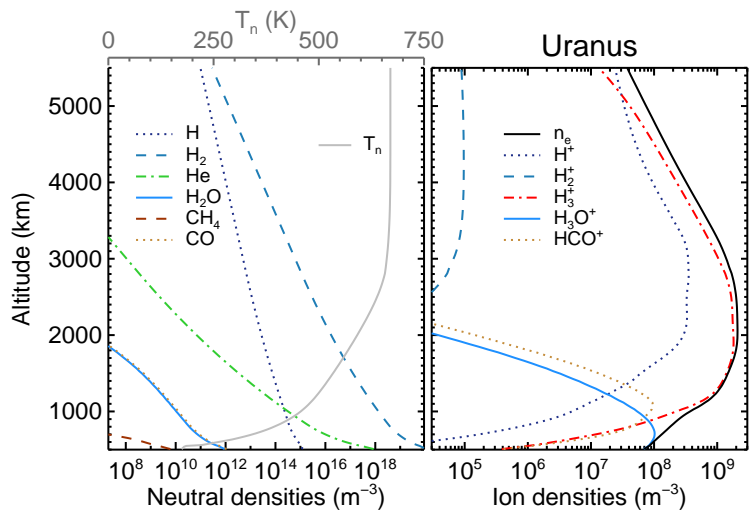}
%%	\adjustimage(width=1.0\textwidth,left,caption={Model profiles for Uranus.  Representative (left) ion density profiles for $30\degree$ S latitude, 12 SLT, and (right) background neutral parameters, which come from \textit{Moses and Poppe }\cite{Moses2017}, and are held fixed at all latitudes.},label={fig:figure-6}, figure={"../../../model runs/Figure 6 - Uranus ions neuts"}
	\caption{Model profiles for Uranus.  Representative \textbf{(left)} background neutral parameters, which come from \textit{Moses and Poppe }\cite{Moses2017} and are held fixed at all latitudes, and \textbf{(right)} ion density profiles for $30\degree$ S latitude, 12 SLT.      }
	\label{fig:figure-6}
\end{figure}

\begin{figure}[hbt!]
	%	\raggedright
	\includegraphics[width=1.1\textwidth,center]{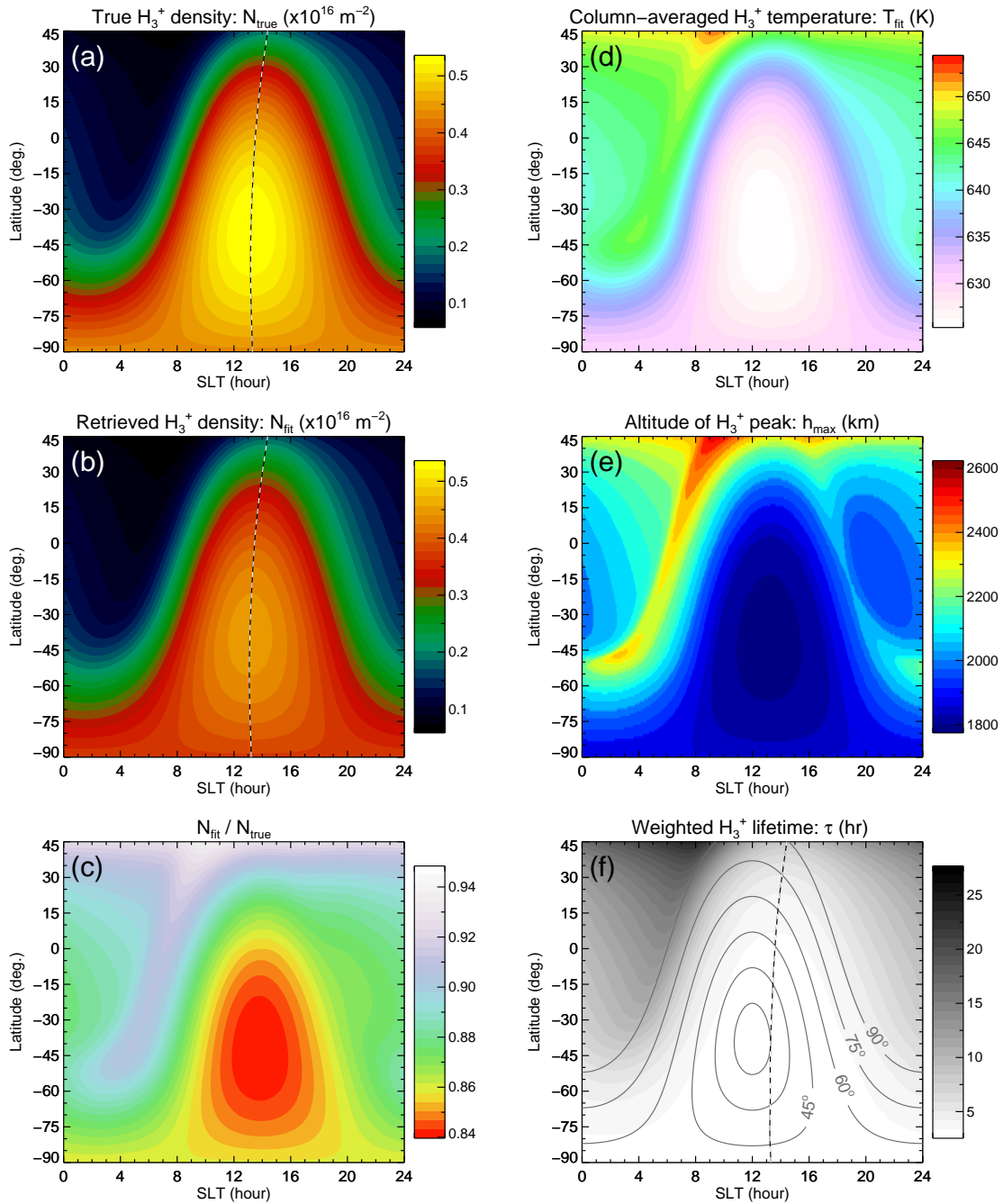}
	\caption{Global ionospheric model results for Uranus.  Contours of \textbf{(a)} the true modelled H$_{3}^{+}$ column density, \textit{N$_{true}$}, \textbf{(b)} the H$_{3}^{+}$ density retrieved from a fit of the modelled spectrum, \textit{N$_{fit}$}, \textbf{(c)} the \textit{N$_{fit}$} / \textit{N$_{true}$} ratio, \textbf{(d)} the column-averaged H$_{3}^{+}$ temperature from a fit to the modelled spectrum, \textbf{(e)} the altitude of the peak of the H$_{3}^{+}$ density, and \textbf{(f)} the column-averaged H$_{3}^{+}$ lifetimes, weighted by H$_{3}^{+}$ density.  In addition, white/black dashed lines in panels (a), (b), and (f) indicate the peak of each parameter in SLT vs. latitude.  Solar zenith angle contours, relevant for all panels, are shown in panel (f).}
	\label{fig:figure-7}
\end{figure}

\label{Effect of vertical atmospheric gradients on H$_{3}^{+}$ observables: Temperature}
While density and temperature structures within a planet's H$_{3}^{+}$ layer affect the retrieved column density, they also affect the interpretation of the retrieved column-averaged temperature.  In order to demonstrate this effect, we model the global distribution of H$_{3}^{+}$ at Uranus.  First, the background atmosphere is based on \textit{Moses and Poppe} \cite{Moses2017}, appropriate for globally-averaged conditions with dust-derived oxygen influxes of 1.2x10$^{5}$ H$_{2}$O molecules cm$^{-2}$ s$^{-1}$, 2.5x10$^{5}$ CO molecules cm$^{-2}$ s$^{-1}$, and 3.0x10$^{3}$ CO$_{2}$ molecules cm$^{-2}$ s$^{-1}$, consistent with H$_{2}$O, CO and CO$_{2}$ observations \cite{Feuchtgruber1997,Cavalie2014a,Orton2014}.  This 1-D atmosphere is applied uniformly at Uranus.  While clearly not fully realistic, using a fixed neutral atmosphere, where ion and neutral chemistry is not fully coupled, allows for clearer elucidation of the effects of a varying H$_{3}^{+}$ layer on retrieved temperatures, as will be demonstrated below.  Next, a simulation date of 15 September 2017 is chosen.  This choice is largely arbitrary for the purposes of demonstrating the effects of gradients on retrieved H$_{3}^{+}$ temperatures; however, there do happen to be contemporaneous H$_{3}^{+}$ observations from September 2017 \cite{Melin2019}, for which Uranus' sub-solar latitude was -$38\degree$.  Finally, 1-D ionospheric calculations are performed globally as described in \ref{Modelling overview}, with a $1\degree$ latitude resolution.  Profiles of background neutral density, temperature, and ion density at $30\degree$ S latitude are shown in Figure \ref{fig:figure-6}.  Electron temperatures (not shown) are calculated to diverge from the neutral temperature around 2000 km altitude and reach \textasciitilde800 K at $30\degree$ S latitude, 12 solar local time (SLT).  (Note that altitude levels throughout the text are referenced to the 1 bar pressure level.)  Calculated H$_{3}^{+}$ temperatures are found to be equal to the background neutral temperature.  
\\ \indent Figure \ref{fig:figure-7} presents global ionospheric density results at Uranus, plotted versus SLT and planetocentric latitude.  Note that, as H$_{3}^{+}$ temperatures were found to be identical to the neutral temperature, at least for the conditions in Figure \ref{fig:figure-6}, these global simulations do not include plasma temperature calculations, and therefore any H$_{3}^{+}$ temperature variations are a reflection of the vertical distribution of H$_{3}^{+}$ and the background temperature profile.  Following the preceding approach, \textit{N$_{fit}$} and \textit{T$_{fit}$} follow from generating and fitting synthetic H$_{3}^{+}$ spectra based on modelled H$_{3}^{+}$ distributions.  These simulations utilize 138 H$_{3}^{+}$ slabs (i.e., grid points in altitude), and therefore \textit{N$_{fit}$} is within \textasciitilde0.2\% of its asymptotic value based on Figure \ref{fig:figure-3}. 
\\ \indent While direct reproduction wasn't the goal of these simulations -- and would be at least partially coincidental  anyway due to the homogenous neutral atmosphere -- calculated column densities are broadly consistent with H$_{3}^{+}$ observations \cite{Melin2013,Melin2019}.  As expected from \ref{Results and discussion}\ref{Effect of vertical atmospheric gradients on H$_{3}^{+}$ observables: Density}, the true modelled H$_{3}^{+}$ column density, panel (a), is slightly larger than the retrieved value, panel (b), with a global mean column density ratio, panel (c), of 0.87.  The column density ratio is nearest to 1.0 at dawn, and at high northern latitudes, indicating that more of the H$_{3}^{+}$ layer there is in the isothermal region of the model atmosphere above 2800 km altitude, as would be expected based on solar zenith angle (SZA) effects and nightside recombination chemistry, which depletes lower-altitude layers near the electron density peak more rapidly \cite{Nagy2009}.  This is more evident from Figure \ref{fig:figure-7}d, which reveals that the column-averaged H$_{3}^{+}$ temperature is higher in those regions, a direct response of the peak altitude of the H$_{3}^{+}$ layer being shifted towards higher altitudes, as seen in Figure \ref{fig:figure-7}e.  This is primarily a SZA effect, as more oblique slant paths through the atmosphere generate higher-altitude photoionization, however there is also a slight offset post-noon due partly to conversion of H$^{+}$ to H$_{2}^{+}$ (and thus, H$_{3}^{+}$, following reaction with H$_{2}$) via reaction \eqref{1.1}.  The H$_{3}^{+}$ lifetime at the H$_{3}^{+}$ peak is on the order of \textasciitilde1 hour for most of the Uranus day, as would be expected based on the modelled peak density (\textasciitilde3000 cm$^{-3}$) and dissociative recombination rate of H$_{3}^{+}$ with electrons (\textasciitilde10$^{-7}$ cm$^{3}$ s$^{-1}$ \cite{Larsson2008}).  Column-averaged H$_{3}^{+}$ lifetimes, weighted by H$_{3}^{+}$ density, are slightly higher than at the peak altitude, but still typically \textless3 hours (Figure \ref{fig:figure-7}f).

Due to the simplified nature of the Uranus simulations, and due to the sparse thermospheric constraints at present, a detailed model-data comparison is not warranted here.  Instead, we emphasize the temperature variations shown in Figure \ref{fig:figure-7}d.  Despite the fact that the thermospheric temperature profile is identical at all latitudes, retrieved H$_{3}^{+}$ temperatures vary by \textgreater35 K, or roughly 5\% of the mean temperature.  This is simply understood as primarily a SZA effect: at low SZAs the H$_{3}^{+}$ layer is lower in the ionosphere, probing the lower temperatures there, whereas the reverse is true for high SZAs.  This result is, again, not surprising; however it highlights two important points: (1) observed H$_{3}^{+}$ temperature variations do not necessarily imply anything about the energetics of the thermosphere, and (2) these SZA effects should be accounted for when interpreting measured temperature variabilities.  Finally, in reviewing the results of Figure \ref{fig:figure-7}, it is important to also emphasize the elements that are missing from the simulations presented there.  In particular, we have neglected energetic particle precipitation, which would be expected to lead to increased ionization and enhanced thermospheric temperatures, mainly at high magnetic latitudes.  At Uranus, the magnetic polar regions are at mid- and low-latitude as a result of the tilted magnetic dipole axis \cite{Herbert2009}, though their exact longitudes are unknown at present due to uncertainty in Uranus' rotation period.  Thus, the chief value of Figure \ref{fig:figure-7} is in demonstrating the qualitative effects of H$_{3}^{+}$ density and temperature gradients on retrieved parameters.  More realistic global variations of ionization and heating at Uranus would be expected to lead to different quantitative structures.

\subsection{Vertical structure of H$_{3}^{+}$ density and temperature (Jupiter)} \label{Vertical structure}
One of the most sought-after observables for planetary atmospheres is the thermal structure, as so much of the rest of planetary dynamics and energetics depend upon it.  Furthermore, as established in sections \ref{Results and discussion}\ref{Effect of vertical atmospheric gradients on H$_{3}^{+}$ observables: Density} and \ref{Results and discussion}\ref{Effect of vertical atmospheric gradients on H$_{3}^{+}$ observables: Temperature}, atmospheric gradients in the H$_{3}^{+}$ layer can confuse measurements of H$_{3}^{+}$ density and temperature, meaning the very parameter we want to constrain, \textit{T(z)}, is itself limiting observational insight.  We now introduce a potential method of unraveling this confusion by combining column-integrated H$_{3}^{+}$ measurements with forward modelling.

\begin{figure}[hbt!]
	\centering
	\includegraphics[width=0.95\linewidth]{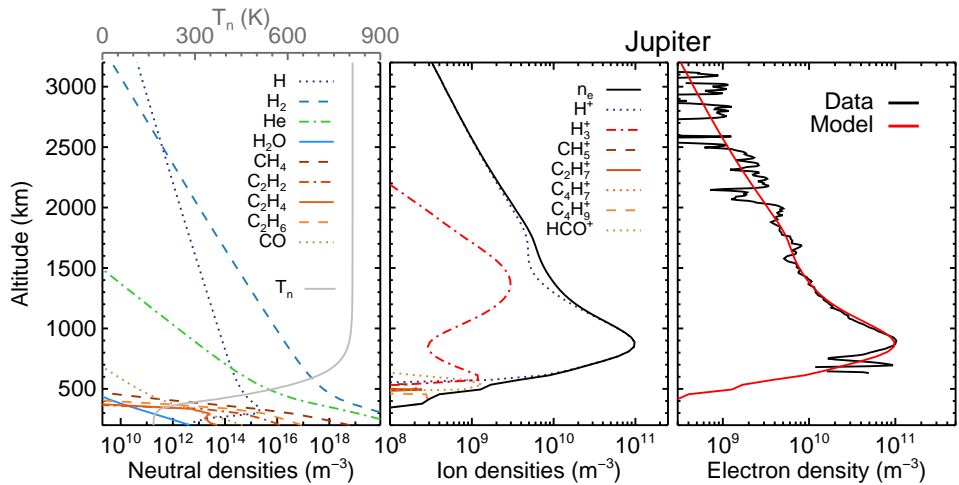}
	\caption{From left to right: the background neutral density and temperature structure; corresponding modelled ion densities; and the Galileo G0N radio occultation electron density profile (black) \cite{Hinson1997} compared with the model reproduction (red).}
	\label{fig:figure-8}
\end{figure}

For this proof-of-concept, we use the Galileo G0N radio occultation, obtained on 8 December 1995, which sampled Jupiter's dusk ionosphere near $24\degree$ S planetocentric latitude and \linebreak $292\degree$ E longitude \cite{Hinson1997}.   Figure \ref{fig:figure-8} presents a model reproduction of the G0N occultation, along with corresponding background neutral atmospheric parameters and modelled ion densities.  The model simulation is for $24\degree$ S latitude, with a forced vertical drift \textit{W$_{D}$} of 50 cm s$^{-1}$, and plasma density comparisons are extracted for 18 SLT in accordance with the dusk terminator measurement of G0N.  As described in \ref{Methods}\ref{Modelling overview}, the solar flux is specified using extrapolated TIMED/SEE measurements, the secondary ionization and photoelectron heating rate are parameterized \cite{Moore2009}, and the effective \eqref{1.1} reaction rate comes from \textit{Majeed et al.}\cite{Majeed1991a}, with the rest of the chemistry as specified in \textit{Moore et al.} \cite{Moore2018a}.

\begin{figure}[hbt!]
	\centering
	\includegraphics[width=0.8\linewidth]{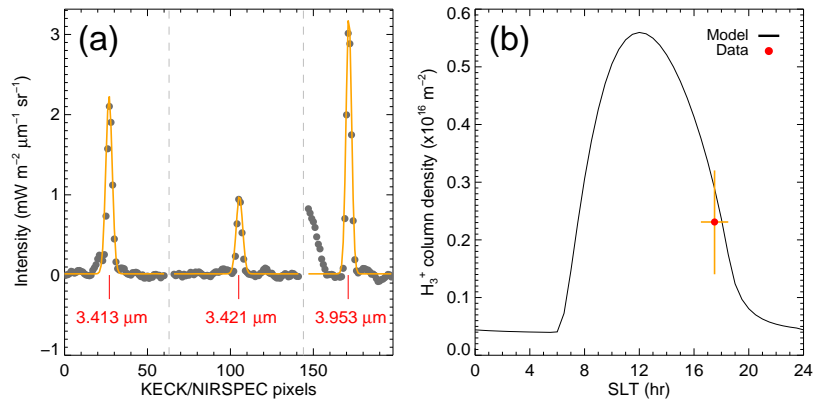}
	\caption{\textbf{(left)} Extracted, calibrated spectral regions from Keck/NIRSPEC observations at Jupiter (grey circles), along with the H$_{3}^{+}$ spectral fit (orange).  \textbf{(right)} Diurnal variation of modelled H$_{3}^{+}$ column density (black), along with the corresponding value retrieved from the Keck observations (red circle).  Vertical error bars come from the spectral fit; horizontal error bars indicate the SLT region over which the spectra were obtained.}
	\label{fig:figure-9}
\end{figure}

The good model-data agreement in Figure \ref{fig:figure-8} gives some confidence that the broad electron density features are well-represented (that is, aside from the noisy densities in the topside and the narrow structures below the peak, which might be attributed to gravity waves \cite{Matcheva2001}).  As an additional test of the model simulation, we turn to H$_{3}^{+}$ observations obtained in April 2016 using Keck/NIRSPEC (section \ref{Methods}\ref{Observations and data reduction}).  First, a series of spectra obtained with the NIRSPEC slit oriented E-W near Jupiter's Great Red Spot are combined and reduced in order to obtain a calibrated H$_{3}^{+}$ spectrum at $24\degree$ S latitude, 17 SLT.  Second, we derive a spectral fit to the data, and compare the fit parameters to the H$_{3}^{+}$ column density from the ionospheric simulation shown in Figure \ref{fig:figure-8}.  An extracted portion of the H$_{3}^{+}$ spectrum, its corresponding spectral fit, and the modelled column densities are shown in Figure \ref{fig:figure-9}.  Retrieved parameters are \textit{T$_{fit}$} = 774 $\pm$ 64 K and \textit{N$_{fit}$} = (2.31 $\pm$ 0.88)x10$^{15}$ m$^{-2}$, and the latter is over plotted on the modelled column densities with horizontal error bars that identify the range of local times that contributed to the observed spectrum.  

There is good model-data agreement in Figure \ref{fig:figure-9}b, which lends additional confidence that the model is well-representing both the observed electron density (Figure \ref{fig:figure-8}) and H$_{3}^{+}$ column density (Figure \ref{fig:figure-9}b).  Before moving on, however, there are a couple of important caveats to emphasize.  First, and perhaps most important, the ionospheric simulation is for 20 April 2016 during solar minimum, in accordance with the Keck H$_{3}^{+}$ observations, whereas the Galileo radio occultation was obtained on 8 December 1995, nearly 21 years prior, also during solar minimum.  It would be far more surprising if Jupiter's ionosphere had not changed in those intervening years than if it had, especially given the variability present in ionospheric radio occultations at giant planets \cite{Yelle2004,Kliore2014}.  Therefore, the fact that the model agreement is good in both Figures \ref{fig:figure-8} and \ref{fig:figure-9} is most likely a coincidence rather than an indication of atmospheric stability, though the fact that the \textasciitilde21 year separation between the two datasets is nearly 2 full solar cycles allows some minimum of hope for a happy coincidence to be maintained.  Second, the Galileo radio occultation was at $68\degree$ W (System III) longitude, whereas the Keck observations were centred at $308\degree$ W longitude, a slightly different magnetic environment.  The majority of the modelled H$_{3}^{+}$ layer is still in photochemical equilibrium, meaning that this variation should have minimal effect, at least if solar photons are the main ionospheric driver as assumed here, though high altitude ion drifts would be altered.  Nevertheless, given that the model is able to reproduce both available datasets, and given that there are no better combinations of ionospheric constraints available, we shall progress forward under the assumption that the ionospheric model simulation provides as accurate a representation of the H$_{3}^{+}$ density structure as possible at present.

To re-state the problem: the observed H$_{3}^{+}$ spectrum is a function of the integrated density\textit{ N(z)} and temperature \textit{T(z) }profiles from the emitting H$_{3}^{+}$ layer.  Thus, the IR spectrum \textit{I($\lambda$)} also contains information about both of them.  Essentially, there are three unknowns, so if either\textit{ N(z)} or \textit{T(z)} can be convincingly constrained then the other can in principle be derived when combined with the observed spectrum.  The spectrum in this example is known from Keck/NIRSPEC observations.  Based on the good model-data agreement for the electron density profile and the H$_{3}^{+}$ column density, we proceed under the assumption that \textit{N(z)} is appropriately constrained.  Therefore, we should also be able to reconstruct at least some limited representation of \textit{T(z)} from the above two inputs.

First, as in section \ref{Results and discussion}\ref{Effect of vertical atmospheric gradients on H$_{3}^{+}$ observables: Density}, the modelled H$_{3}^{+}$ layer is idealized as \textit{n} slabs, each of column density \textit{N$_{slab}$ = N$_{true}$ / n}.  Second, a temperature is assigned to each slab, randomly selected from \linebreak 150-1250 K, and a synthetic H$_{3}^{+}$ spectrum (i.e., the sum of the slab spectra) is computed.  In principle, the specified temperatures of each slab could be independent of each other, but for the present case a monotonically increasing (or isothermal) temperature profile is enforced, consistent with observation \cite{Yelle2004,Lystrup2008}.  This step is repeated until the modelled and observed spectra converge to within some pre-defined tolerance.  In this case the tolerance is set to a maximum of 0.03\% disagreement between the slab-modelled spectrum and the spectrum obtained from the H$_{3}^{+}$ fit.  Once converged, it is useful to also provide some estimate of the sensitivity of the result.  For this purpose, \textit{n-1} slab temperatures are held fixed to their converged values while the other is freely varied until the modelled column-averaged temperature exceeds the measured temperature uncertainty.  The derived temperature uncertainties thus represent \textasciitilde8\% errors in this case, as \textit{T$_{fit}$} = 774 $\pm$ 64 K.  Finally, based on the converged slab temperatures weighted by their uncertainties, an analytical temperature profile is derived following \cite{Krasnopolsky2002}:
\begin{equation}\label{1.2}
%	\begin{split}
		T(z) = T_{exo} - (T_{exo} - A_1)\ exp\,\left[-\frac {(z-A_2)^2}{A_3T_{exo}}\right]
%	\end{split}
\end{equation}
where \textit{z} is the altitude element, \textit{T$_{exo}$ }the exospheric temperature, and \textit{A$_{i}$} are constants of the fit.

\begin{figure}[hbt!]
	\centering
	\includegraphics[width=0.7\linewidth]{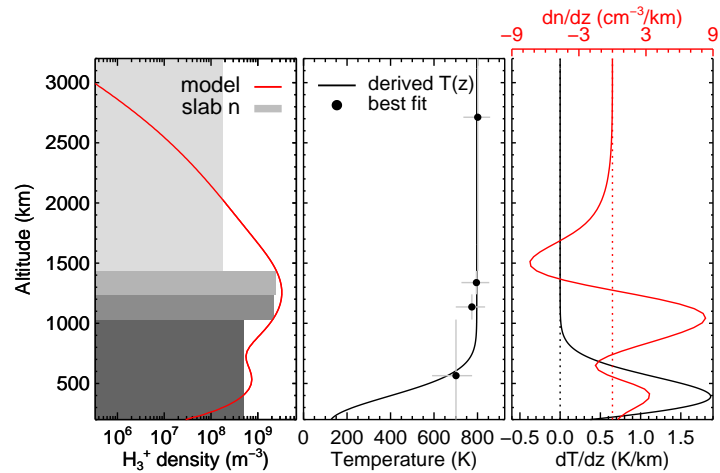}
	\caption{\textbf{(left)}The modelled H$_{3}^{+}$ density (red), represented as 4 slabs of equal column density (and thus varying vertical extent and number density; grey shading).  \textbf{(middle)} Best-fit slab temperatures (black circles), along with estimated uncertainties (grey lines), and the derived temperature profile (black).  \textbf{(right)} Corresponding density (red) and temperature (black) gradients.  See text for description of methods.}
	\label{fig:figure-10}
\end{figure}

Results from the above process with \textit{n} = 4 are shown in Figure \ref{fig:figure-10}.  Each slab is represented by a shaded column, with varying vertical extent due to enforced equal slab column densities and variable slab number densities.  The original modelled H$_{3}^{+}$ density is given by the red curve.  These variable slab widths enable higher altitude resolution near the H$_{3}^{+}$ density peak and allow for more reasonable error bars than significantly smaller slab widths would.  Corresponding slab temperatures are shown as black circles on the right, along with the derived temperature profile in red.  Horizontal grey error bars indicate the 1$\sigma$ uncertainty in slab temperature, and vertical grey error bars demarcate the vertical extent of each slab.  The best-fit parameters for the temperature profile in Figure \ref{fig:figure-10} are \textit{T$_{exo}$} = 788 K, \textit{A$_{1}$} = 136 K, \textit{A$_{2}$} = 159 km, and \textit{A$_{3}$} = 121 km$^{2}$ K$^{-1}$, and are only representative of the altitude range with significant H$_{3}^{+}$ density (e.g., between \textasciitilde300-2500 km).

Combining the results of Figure \ref{fig:figure-10} with those from section \ref{Results and discussion}\ref{Effect of vertical atmospheric gradients on H$_{3}^{+}$ observables: Density}, it appears that the Keck/NIRSPEC observations were minimally affected by gradients in the H$_{3}^{+}$ layer.  The derived thermal gradient is effectively zero above 1050 km (i.e., $\le$0.05 K/km), $\textless$0.2 K/km for altitudes between 750-1050 km, \textasciitilde0.3-1.4 between 550-750 km, and \textasciitilde1.8 K/km at the bottom side.  In situ measurements at Jupiter \cite{Seiff1998} and Saturn \cite{Yelle2018} find \textasciitilde2 K/km near 400 km and 0.4 K/km at the base of the thermosphere, respectively.  Based on Figure \ref{fig:figure-5}, this implies that retrieved H$_{3}^{+}$ column densities are less than 80\% of the modelled values only for altitudes $\textless$750 km, a range that encompasses 12\% of the total column density.  Meanwhile, absolute modelled density gradients are $\textless$8 cm$^{-3}$/km everywhere, and only \textasciitilde3 cm$^{-3}$/km where there is also a substantial temperature gradient (near 400 km altitude).  Progression of the impact of the derived density and temperature gradients on retrieved column density is shown by the dashed gray curve in Figure \ref{fig:figure-5}b.  In total, the calculated error in column density, based on the combination of \ref{fig:figure-5}b and the density structure from Figure \ref{fig:figure-10}, is \textasciitilde10\%, all associated with gradients at altitudes below 800 km.  This error is well-within the observational uncertainties (Figure \ref{fig:figure-9}).

\section{Conclusions}
This study has investigated the effect of atmospheric H$_{3}^{+}$ density and temperature gradients on the interpretation of observations of the composite spectrum.  Overall atmospheric structure is found to cause observed H$_{3}^{+}$ column densities, \textit{N$_{fit}$}, to represent a lower limit.  The degree to which \textit{N$_{fit}$} constrains the true ionospheric column density, \textit{N$_{true}$}, depends primarily on the magnitude of any temperature gradients and secondarily on the mean temperature within the H$_{3}^{+}$ layer.  Density gradients can act to reduce the retrieved \textit{N$_{fit}$} / \textit{N$_{true}$} ratio even further, provided there is also a temperature gradient present.

Giant planets generally exhibit strong positive temperature gradients in the lower thermosphere, and consequently low altitude H$_{3}^{+}$ is most significantly underestimated.  This is also the region where a majority of H$_{3}^{+}$ is produced, and where the atmosphere is most electrically conductive.  Therefore, one immediate caution based on the above results is that nearly all of the error in retrieved H$_{3}^{+}$ densities due to atmospheric density and temperature gradients is associated with this low altitude region.  The total error in H$_{3}^{+}$ column density from nadir column-integrated observations may be small (e.g., 10\%), but the local error in H$_{3}^{+}$ number density can be large (typically 50\% or more, Figure \ref{fig:figure-5}), and this should be considered when estimating ionospheric electrical conductivities associated with H$_{3}^{+}$. 

Based on thermal structure in the atmosphere, derived H$_{3}^{+}$ temperatures are found to represent primarily the temperature at the H$_{3}^{+}$ density peak.  This result, while not surprising, does serve to emphasize that observed H$_{3}^{+}$ temperature variations may not always represent any inherent evolution in the thermosphere, and may instead be attributed to simple photochemical effects.  For example, an X-ray flare would lead to a brief burst of high energy photons, producing a low altitude ionization layer and weighting derived H$_{3}^{+}$ temperatures towards the (generally) lower temperatures there.  Similarly, high solar zenith angle regions will absorb ionizing radiation higher in the atmosphere and therefore exhibit higher average H$_{3}^{+}$ temperatures.  This same effect is expected post-sunset, as dissociative recombination with electrons more quickly depletes the lower altitude H$_{3}^{+}$ near the electron density peak, shifting the effective H$_{3}^{+}$ layer towards higher and higher altitude throughout the night.

While this work gives some guidance on the degree to which atmospheric gradients induce additional uncertainty in retrieved column-integrated H$_{3}^{+}$ densities and temperatures, real-world atmospheric variations far-outstrip those considered here.  Therefore, these calculations cannot represent a definitive quantitative manual.  Instead, their primary value is in providing qualitative insight, and in demonstrating the potential for enhancing the scientific impact of H$_{3}^{+}$ observations through complementary modelling studies. 

Finally, by combining two ionospheric data sets with a model simulation that accounts for the above effects, a method for deriving a temperature profile from an overhead H$_{3}^{+}$ observation is presented.  This method relies on the relatively straightforward nature of H$_{3}^{+}$ solar-driven photochemistry at non-auroral latitudes, and furthermore requires at least some knowledge of the atmospheric structure, so is not easily applicable everywhere.  Nevertheless, given the abundance of H$_{3}^{+}$ observations already obtained, especially at Jupiter and Uranus, it offers potential for improving constraints on global temperature variations at those planets, and it serves as a first-step towards developing a complete H$_{3}^{+}$ retrieval tool.  Ideally, an independent, more traditional method of deriving thermal profiles could first be used to validate this approach.  The H$_{3}^{+}$ limb profiles obtained by the JIRAM instrument \cite{Adriani2014} on-board the Juno spacecraft at Jupiter may represent the perfect opportunity for such a validation.

%%\section{Figures \& Tables}

%%\begin{figure}[!h]
%%\centering\includegraphics[width=2.5in]{test.eps}
%%% where xxxxxx name represents "figurename.eps"
%%\caption{Insert figure caption here}
%%\label{fig_sim}
%%\end{figure}

%%\section{Conclusion}
%%The conclusion text goes here.
\vskip6pt

\enlargethispage{20pt}

%%\ethics{Insert ethics statement here if applicable.}

\dataccess{The data used herein is available in the public archives or from LM upon request.}

\aucontribute{LM led the project, contributed to collection, reduction and analysis of ground-based data, performed ionospheric modelling simulations, produced the figures, and wrote the paper.  HM contributed to collection, reduction and analysis of ground-based data, provided routines for generating synthetic H$_{3}^{+}$ spectra, and took part in detailed discussions.  JO'D wrote the Keck observing proposal, contributed to collection, reduction and analysis of ground-based data, provided inputs and advice for interpreting and generating H$_{3}^{+}$ spectra, and took part in detailed discussions.  TS contributed to collection, reduction and analysis of ground-based data, including specialized spectral fitting techniques, and took part in detailed discussions.  JM conducted neutral atmospheric modelling simulations, and provided detailed guidance in incorporating those results in ionospheric calculations.  MG contributed to development of the secondary ionization and thermal electron heating parameterizations in the modelling.  SM took part in detailed discussion, and provided expert guidance on the history, chemistry and spectral behaviour of H$_{3}^{+}$.  CS contributed to discussion of atmospheric retrieval methods, sensitivities, and visualization of results.  All authors reviewed and edited the manuscript.}

\competing{We have no competing interests.}

\funding{LM was supported by the National Aeronautics and Space Administration (NASA) under Grant NNX17AF14G issued through the SSO Planetary Astronomy Program and Grant 80NSSC19K0546 issued through the Solar System Workings Program.  JM acknowledges support from NASA Solar System Workings grants NNX16AG10G and 80NSSC19K0546. MG acknowledges support from STFC of UK under grant ST/N000692/1.}

\ack{Some of the data presented herein were obtained at the W. M. Keck Observatory, which is operated as a scientific partnership among the California Institute of Technology, the University of California and the National Aeronautics and Space Administration. The Observatory was made possible by the generous financial support of the W. M. Keck Foundation. We are grateful to the TIMED/SEE PI, Tom Woods, and his team for providing us with the solar flux data set and associated routines for extrapolation to planets.}

%%\disclaimer{Insert disclaimer text here if applicable.}

%%%%%%%%%% Insert bibliography here %%%%%%%%%%%%%%

%%\raggedright
\bibliographystyle{vancouver}
%%\bibliographystyle{nature}
%%\bibliography{library}
\bibliography{../../../../../library}
%%\begin{thebibliography}{9}

%%\bibitem{1} Allwood JM, Cullen JM. 2011 \textit{Sustainable materials:  with both eyes open}.
%%Cambridge, UK: UIT Cambridge. See \href{http://www.withbotheyesopen.com}{http://www.withbotheyesopen.com}.

%%\bibitem{2}  MacKay DJC. 2008  \textit{Sustainable energy:  without the hot air}.
%% Cambridge, UK: UIT Cambridge. See \href{http://www.withouthotair.com}{http://www.withouthotair.com}.

%%\bibitem{3} Gallman PG. 2011  \textit{Green alternatives and national energy strategy: the facts
%% behind the headlines}.  Baltimore,\ MD: Johns Hopkins University Press.

%%\bibitem{4} MacKay DJC. 2013.  Solar energy in the context of energy use, energy transportation, and
%% energy storage. \textit{Proc. R. Soc. A} \textbf{371}.

%%\end{thebibliography}

\end{document}